\documentclass{jfmtmp}

\usepackage{natbib}
\usepackage{graphicx}
\usepackage{bm}
\usepackage[tight,hang,it]{subfigure}

\ifCUPmtlplainloaded \else
  \checkfont{eurm10}
  \iffontfound
    \IfFileExists{upmath.sty}
      {\typeout{^^JFound AMS Euler Roman fonts on the system,
                   using the 'upmath' package.^^J}%
       \usepackage{upmath}}
      {\typeout{^^JFound AMS Euler Roman fonts on the system, but you
                   dont seem to have the}%
       \typeout{'upmath' package installed. JFM.cls can take advantage
                 of these fonts,^^Jif you use 'upmath' package.^^J}%
      }
  \else
  \fi
\fi

\ifCUPmtlplainloaded \else
  \checkfont{msam10}
  \iffontfound
    \IfFileExists{amssymb.sty}
      {\typeout{^^JFound AMS Symbol fonts on the system, using the
                'amssymb' package.^^J}%
       \usepackage{amssymb}%
       \let\le=\leqslant  
       \let\ge=\geqslant  
      }{}
  \fi
\fi

\ifCUPmtlplainloaded \else
  \IfFileExists{amsbsy.sty}
    {\typeout{^^JFound the 'amsbsy' package on the system, using it.^^J}%
     \usepackage{amsbsy}}
    {}
\fi

\providecommand\udelta{\delta}

\newcommand\Rey{\mbox{\textit{Re}}}  % Reynolds number
\newcommand\Stok{\mbox{\textit{St}}} % Stokes number
\newcommand\e{\mbox{e}} % Exponential
\newcommand\X{{\bf X}}
\newcommand\V{{\bf V}}
\newcommand\R{{\bf R}}

\newcommand\etal{\mbox{\textit{et al.}}}
\newcommand\etc{etc.\ }

\title[Multifractal concentrations of inertial particles]
        {Multifractal concentrations of inertial particles in smooth
        random flows}

\author[J. Bec]%
{J\ls \'{E}\ls R\ls \'{E}\ls M\ls I\ls E\ns B\ls E\ls C}

\affiliation{E-mail: jeremie.bec@roma1.infn.it\\[\affilskip]
  Dipartimento di Fisica, Universit\`{a} La Sapienza, P.zzle Aldo Moro
  2, 00185 Roma, Italy.\\[\affilskip] D\'epartement Cassiop\'ee,
  Observatoire de la C\^{o}te d'Azur, BP4229, 06304 Nice Cedex 4,
  France.}

\date{16 February 2004}

\pubyear{20??}
\volume{???}
\pagerange{???--???}

\begin{document}

%\addtolength{\baselineskip}{10pt}

\maketitle

\begin{abstract}
  \noindent Collisionless suspensions of inertial particles
  (finite-size impurities) are studied in two- and three-dimensional
  spatially smooth flows. Tools borrowed from the study of random
  dynamical systems are used to identify and to characterise in full
  generality the mechanisms leading to the formation of strong
  inhomogeneities in the particle concentration.
  
  Phenomenological arguments are used to show that in two dimensions,
  the positions of heavy particles form dynamical fractal clusters
  when their Stokes number (non-dimensional viscous friction time) is
  below some critical value.  Numerical simulations provide strong
  evidence for the presence of this threshold in both two and three
  dimensions and for particles not only heavier but also lighter than
  the carrier fluid.  In two dimensions, light particles are found to
  cluster at discrete (time-dependent) positions and velocities in
  some range of the dynamical parameters (the Stokes number and the
  mass density ratio between fluid and particles). This regime is
  absent in the three-dimensional case for which evidence is that the
  (Hausdorff) fractal dimension of clusters in phase space
  (position--velocity) remains always above two.
  
  After relaxation of transients, the phase-space density of particles
  becomes a singular random measure with non-trivial multiscaling
  properties, whose exponents cannot be predicted by dimensional
  analysis.  Theoretical results about the projection of fractal sets
  are used to relate the distribution in phase space to the
  distribution of the particle positions.  Multifractality in phase
  space implies also multiscaling of the spatial distribution of the
  mass of particles. Two-dimensional simulations, using simple random
  flows and heavy particles, allow the accurate determination of the
  scaling exponents: anomalous deviations from self-similar scaling
  are already observed for Stokes numbers as small as $10^{-4}$.
  
\end{abstract}

\section{Introduction}
\label{sec:intro}

Suspensions of dust, droplets, bubbles or other kind of small
particles in turbulent incompressible flows are present in many
stirring and mixing problems encountered in both natural and
industrial situations.  Such impurities typically have a finite size
and a mass density different from that of the carrier fluid.  They
cannot be described as simple passive tracers, that is point-like
particles with negligible mass advected by the fluid; an accurate
model for their motion must take into account inertia effects due to
the finiteness of their sizes and masses.  These \emph{inertial
particles} interact with the fluid through a viscous Stokes drag and
thus their motion typically lags behind that of passive tracer
particles.  The dynamics of the latter is governed by a conservative
dynamical system when the carrier flow is incompressible (because
volume is conserved), but inertial particles have, as we shall see,
\emph{dissipative} dynamics.  While an initially uniform
distribution of passive tracers remains uniform at any later time, the
spatial distribution of inertial particles develops strong
inhomogeneities.

Such a phenomenon, frequently referred to as \emph{preferential
  concentration}, is associated with the presence of regions with
  either extremely high or low concentrations.  Their characterisation
  plays an essential role in the understanding of natural and
  industrial phenomena.  Instances are optimisation of combustion
  processes in the design of Diesel engines \cite[]{ekllpr03}, the
  growth of rain drops in subtropical clouds \cite[]{pk97,ffs02}, the
  appearance of planetesimals in the early stage of planet formation
  in the Solar system \cite[]{w95,chpd01}, coexistence problems
  between several species of plankton \cite[]{sy95,kpstt00} and other
  environmental problems \cite[see, e.g.,][]{s86}.  For such
  applications it is recognised that a key problem is the prediction
  of the collision or reaction rates and their associated typical time
  scales.  The traditional way to estimate the latter goes back to
  \cite{t21} and makes use of diffusion theory.  The time scales
  obtained in such approaches generally exceed by one or several
  orders of magnitude those observed in experiments or numerical
  simulations \cite[see, e.g.,][]{sc97}. A full understanding of
  particle clustering and, in particular, of the fine structures
  appearing in the mass distribution is crucial for identifying and
  quantifying the mechanisms responsible for the drastic reduction in
  time scales.

We propose in this paper an original approach leading to a systematic
description of inertial particles clustering.  This approach is in
part inspired by recent major breakthroughs in the study of passive
scalar advection by turbulent flows, using Lagrangian techniques
\cite[see, e.g., the review of ][]{fgv01}.  Preferential
concentrations are due to the convergence of inertial particle
trajectories onto certain sets in the position--velocity phase space
called \emph{attractors}, which are usually fractals.  Of course,
these sets are dynamically evolving objects and depend on the history
of the carrier flow.  When interested only in the spatial distribution
of particles, one clearly has to consider the projection on positions
of the phase-space attractor. Projection of singular sets are
themselves generally singular. As a consequence, an initially uniform
distribution of particles will tend to become singular at large times,
after relaxation of transients. This is the basic mechanism
responsible for the formation of clusters of particles.  Use of
dissipative dynamical systems tools and, in particular, of methods
borrowed from the study of random dynamics, allows a rather complete
characterisation of the particle distribution at those scales where
the carrier flow can be considered smooth in space.  As we shall see,
the statistical properties of the spatial distribution of particles in
these clusters can be characterised as a function of the dynamical
parameters (which depend on the physical properties of the fluid and
on the masses and sizes of the particles).

We focus on a simple model which captures most qualitative aspects of
inertial particles.  In this simplified dynamics, only two effects are
included: the Stokes drag proportional to the velocity difference
between the particle and the fluid and an added mass term proportional
to the acceleration of the fluid volume displaced by the particle.
These two terms are associated to two parameters: the Stokes number
$\Stok$ proportional to the cross section of the particle and the mass
density ratio between the particle and the surrounding fluid.  We
identify and characterise the different regimes of the dynamics
appearing when varying these two parameters.

The paper is organised as follows. In \S\ref{sec:equations} we derive
a simple model for the dynamics as an approximation of the full Newton
equation describing the motion of particles; in particular, we justify
its relevance in some asymptotic range of the physical parameters of
the problem.  In \S\ref{sec:local} this model is used to relate the
local dynamics of the particles to the local properties of the carrier
flow; this permits to perform a complete stability analysis of the
particle dynamics.  In \S\ref{sec:threshold}, after a brief
introduction to the tools used in studies of dissipative dynamics with
particular emphasis on Lyapunov exponents (\S\ref{subsec:diss-lyap}),
phenomenological arguments are used to extend the local approach to
the global dynamics in order to show that below a critical value of
the Stokes number the particles concentrate on fractal dynamical
clusters in the position space (\S\ref{subsec:heuristics}); this is
confirmed by numerical experiments presented in
\S\ref{subsec:numerics}. A more precise description of the statistical
properties of these clusters is performed in \S\ref{sec:mass} where it
is shown that the distribution of particle actually has multifractal
properties, even at very small Stokes numbers.  Finally,
\S\ref{sec:conclusion} encompasses concluding remarks and open
questions.

\section{Dynamical model for the particles}
\label{sec:equations}

We consider very diluted suspensions of inertial particles where
collisions, particle-to-particle hydrodynamical interactions and
retroaction of the particles on the motion of the fluid can be
ignored.  It is assumed that the carrier flow has moderate Reynolds
numbers and that the particle radius is smaller than their dissipation
scale. Under the additional assumption that the Reynolds number based
on the particle size and its relative velocity with respect to the
fluid is sufficiently small, \cite{mr83} proposed approaching the flow
surrounding the small sphere by a Stokes flow. Neglecting both the
effects of gravity and the Fax{\'{e}}n corrections, the motion $\X(t)$
of an inertial particle is then solution to the following Newton
equation
\begin{eqnarray}
    m_p \ddot\X &=& m_f \frac{{\rm D}\bm u}{{\rm D}t}(\X, t)
    -6\pi a\mu \left[\dot\X - \bm u(\X, t) \right ]
    \nonumber -\frac{m_f}{2} \left[ \ddot\X - \frac{\rm d}{{\rm
    d}t} \left( \bm u(\X, t) \right)\right ] \nonumber \\ && -
    \frac{6\pi a^2\mu }{\sqrt{\pi \nu}} \int_{0}^{t} \frac{{\rm
    d}s}{\sqrt{t-s}}\, \frac{\rm d}{{\rm d}s}\left[ \dot\X - \bm
    u(\X, s) \right ]. \label{eq:maxeyandriley}
\end{eqnarray}
The dots denote time derivatives, $\bm u$ is the velocity field of the
carrier fluid, ${\rm D}\bm u/{\rm D}t$ is its derivative along the
path of a fluid element, $m_p$ is the mass of the particle, $m_f$ is
the mass of fluid displaced by the particle, $a$ is the radius of the
particle and, finally, $\mu$ and $\nu$ are respectively the dynamic
and the kinematic viscosities of the fluid.  The different forces
exerted on the particle are, in order of appearance on the right-hand
side of (\ref{eq:maxeyandriley}), the force exerted by the undisturbed
flow, the Stokes viscous drag, the added mass effect and the
Basset--Boussinesq history force.

For rescaling purposes it is more convenient to write the equation
satisfied by the velocity difference between the particle and the
carrier fluid $\bm w(t) \equiv \dot\X - \bm u(\X,t)$. Indeed, the
dynamics involves two velocity scales which are determined by $\bm u$
and $\bm w$ and there is \emph{a priori} no reason why they should be
of the same order of magnitude.  Rescaling of space, time, velocity of
the carrier fluid and velocity difference by the factors $L$, $L/U$,
$U$ and $W$, respectively, leads to
\begin{eqnarray}
    \frac{{\rm d}\bm w}{{\rm d}t} = \frac{\beta-1}{\alpha} \frac{\rm
    d}{{\rm d}t} (\bm u(\X, t)) - \beta \left(\bm w\cdot \nabla
    \right) \bm u(\X,t) - \frac{1}{\Stok}\,\bm w
    -\sqrt{\frac{3\beta}{\pi\, \Stok}} \int_0^t \frac{{\rm
    d}s}{\sqrt{t-s}} \,\frac{{\rm d}\bm w}{{\rm d}s}.
    \label{eq:maxeyandrileyadim}
\end{eqnarray}
The dynamics depends here on three parameters\,: the velocity ratio
$\alpha\equiv W/U$, the added-mass factor $\beta \equiv
3m_f/(m_f+2m_p)$ and the \emph{Stokes number} associated to the
particle $\Stok \equiv (a^2 U)/(3\beta \nu L)$.  The latter can be
also written as $\Stok = \Rey \,(a/L)^2 /(3\beta)$ where $\Rey\equiv
UL/\nu$ is the Reynolds number of the carrier flow. The particle
radius $a$ is taken to be smaller than the typical (integral) scale
$L$ of the carrier flow. As a consequence, if the Reynolds number is
not very large and the particles are not very heavy we must assume
that the Stokes number is much smaller than unity.  Moreover, the
requirement of small local Reynolds number in the neighbourhood of the
particle - required in Maxey \& Riley derivation of
(\ref{eq:maxeyandriley}) - reads here $Wa/\nu\ll 1$.  It restricts the
range of admissible parameters to satisfying $\alpha \ll (\beta
\,\Stok)^{-1}$.  When both $\beta$ and $\Stok$ are taken order unity,
the velocity ratio $\alpha$ must be small compared to unity, meaning
that the inertial-particle dynamics must be very close to that of
ordinary fluid particles.  This can only be achieved by requiring low
values of the Stokes number.  Hence, in the Maxey \& Riley approach,
$\beta$ and $\Stok$ cannot be simultaneously order unity.

We nevertheless consider here particles whose motion is described by a
simplified dynamics where both the second and the last term of
(\ref{eq:maxeyandrileyadim}) are neglected.  This is clearly the
case when the typical velocity difference $\alpha$, and thus the
Stokes number are both very small.  Such  simplified dynamics is also
relevant when the particles are much heavier than the fluid
($\beta\ll1$), a   case  of importance for applications involving
suspensions of liquid droplets or dust particles in a gas. Under one
of these two hypotheses and by introducing what we call here 
the \emph{covelocity}
$\V \equiv \dot\X - \beta\,\bm u (\X,t)$, it is easy to
see that the equation of motion reduces to the $(2\times
d)$-dimensional system
\begin{subeqnarray}
  \dot\X &=& \beta\,\bm u (\X,t) + \V, \\ \dot\V
  &=& \frac{1}{\Stok} \left[ (1-\beta)\, \bm u(\X,t) - \V
  \right], \label{eq:newton}
\end{subeqnarray}
combined with the initial conditions $\X(0)=\bm x_0$ and $\V(0) =
(1-\beta)\,\bm u({\bm x}_0,0) + \alpha\bm w_0$. Here $\bm w_0$ denotes
the non-dimensional initial velocity difference between the particle
and the flow.  The dynamics described by (\ref{eq:newton}) are clearly
dissipative.  Indeed, the divergence of the right-hand side of
(\ref{eq:newton}) with respect to the phase-space variables $(\bm x,
\bm v)$ reduces in the case of divergence-free carrier flow to $-
d/\Stok$, which is negative. This implies that all volumes in
position--covelocity phase space are uniformly contracted during time
evolution.

\section{Local analysis of the dynamics}
\label{sec:local}

As a first step in describing the mechanisms responsible of particle
clustering, we study the time evolution of the position--covelocity
phase-space separation $\R \equiv (\udelta\X, \udelta\V)$ between two
infinitesimally close trajectories.  Since the sizes of inertial
particles are within the dissipation range of the carrier flow, the
velocity field $\bm u$ can be considered spatially smooth.  Velocity
gradients are then bounded and velocity differences between two
neighbouring locations are linearly proportional to their
separation. The separation $\R(t)$ thus obeys an equation obtained by
linearising (\ref{eq:newton}), namely
\begin{equation}
\dot\R = {\cal M}_t \,\R, \qquad {\cal M}_t\equiv \left[
\begin{array}{cc} \beta \,\bm \sigma(t) & {\cal I}_d \\[5pt]
\displaystyle\frac{1-\beta}{\Stok}\, \bm \sigma(t) &
\displaystyle-\frac{1}{\Stok}\, {\cal I}_d \end{array} \right].
\label{eq:tangentsys}
\end{equation}
Here $\bm \sigma$ denotes the strain matrix of the carrier flow along
the trajectory of a particle: $\sigma_{ij}(t) \equiv \partial_j u_i
({\bm X}(t), t)$, and ${\cal I}_d$ stands for the $d$-dimensional
identity matrix. Qualitative insight into the local dynamics of the
particles is provided by making the stability analysis of
(\ref{eq:tangentsys}).  The eigenvalues of the evolution matrix ${\cal
M}_t$ depend on the local structure of the carrier flow, as we now
show.  It is easily checked that ${\cal M}_t$ is a solution to the
second-order equation
\begin{equation}
  {\cal M}_t\!^2 + \left( \frac{1}{\Stok}\, {\cal I}_{2d} - \beta\,
  \Sigma_t \right) {\cal M}_t - \frac{1}{\Stok}\, \Sigma_t =
  0_{2d},
  \label{eq:2ndorderM}
\end{equation}
where $\Sigma_t$ is the $2d\times2d$ block-diagonal matrix given by
\begin{equation}
  \Sigma_t \equiv \left[ \begin{array}{lr} \bm\sigma(t) & 0_d \\ 0_d &
    \bm\sigma(t) \end{array} \right],
  \label{eq:defSigma}
\end{equation}
and $0_d$ is the $d$-dimensional square matrix with all entries zero.  A
straightforward consequence is that, to a given eigenvalue $\gamma$ of
the strain matrix, correspond two eigenvalues of the evolution matrix
${\cal M}_t$ solutions of the quadratic equation
\begin{equation}
  x^2 + \left(\frac{1}{\Stok} - \beta\,\gamma\right) x -
  \frac{\gamma}{\Stok} = 0.
  \label{eq:2ndordervp}
\end{equation}
The stability analysis of the evolution matrix requires to express the
sign of the real part of the solutions to this second-order equation
as a function of $\gamma$, and thus depends on the local structure of
the carrier flow.

Let us examine the consequence, first in two dimensions.  Depending on
the sign of the Okubo--Weiss parameter $Q$ \cite[]{o70,w91} defined as
the determinant of the strain matrix $\bm\sigma$ and thus satisfying
$\gamma^2=Q$, the incompressible flow of the carrier fluid separates
the space in regions of two kinds: the hyperbolic regions where $Q>0$
and where the strain matrix $\bm \sigma$ has two opposite real
eigenvalues and the elliptic regions where $Q<0$, so that the two
eigenvalues are imaginary numbers. A simple analysis of
(\ref{eq:2ndordervp}) shows that in the hyperbolic regions, the
evolution matrix ${\cal M}_t$ of the particle dynamics has three
stable and one unstable eigendirections in the four-dimensional
position--covelocity phase space. In the elliptic regions, one must
distinguish two cases: if $\beta<1$ (particles heavier than the
fluid), there are two stable and two unstable eigendirections, whereas
if $\beta>1$ (lighter particles), there are always four stable
directions. This local analysis suggests that in two dimensions, heavy
particles are excluded from the vortices (elliptic regions) and tend
to concentrate within the filaments (hyperbolic regions) of the
carrier flow, whereas light particles tend to cluster in the vortices.

\begin{figure}
\centerline{\includegraphics[width=0.8\textwidth]{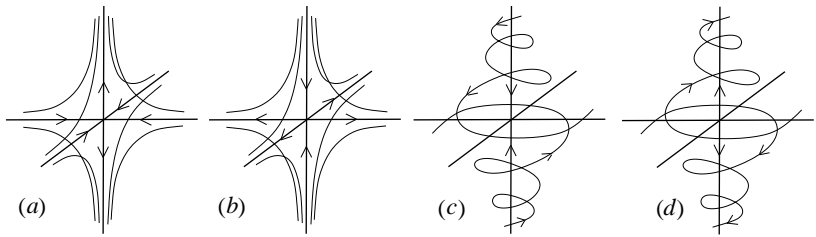}}
\caption{\label{fig:zones3d} Local structure of the carrier flow in
  three dimensions\,; four cases are distinguished accordingly to the
  nature of the three eigenvalues of the strain matrix: if all three
  are real (hyperbolic case), either two of them are negative and one
  is positive (\textit{a}) or two of them are positive and the third
  is negative (\textit{b}). When two eigenvalues are complex conjugate
  (elliptic case), the third one is real and can be either negative
  (\textit{c}) or positive (\textit{d}).}
\end{figure}
In three dimensions, there are four possible local structures of the
incompressible carrier flow represented schematically in
figure~\ref{fig:zones3d}. They can also be divided into elliptic and
hyperbolic cases depending on the presence or not of complex
eigenvalues.
\begin{itemize}
\item \emph{Hyperbolic regions} (\textit{a}, \textit{b}): the local
  dynamics of the particles has five stable and one unstable
  eigendirections in the case (\textit{a}) and four stable and two
  unstable in the case (\textit{b});
\item \emph{Elliptic regions} (\textit{c}, \textit{d}): there is
  generally at least one unstable eigendirection; the exception is
  case (\textit{c}) for light particles ($\beta>1$); all six
  eigendirections are then stable provided the real negative
  eigenvalue $\gamma$ of the strain matrix satisfies
  $\gamma<-\Stok/\beta$.
\end{itemize}

The three-dimensional situation looks hence similar to the
two-dimensional one: the local analysis suggests that particles
lighter than the carrier fluid concentrate in the rotation regions
while heavier particles tend to be ejected from them and concentrate
in the strain regions. The model is hence able to catch a qualitative
property of inertial particles dynamics which plays a central role in
the phenomenological understanding of their preferential
concentration.  As stated for instance in \cite{se91}, the appearance
of inhomogeneities in the distribution of particles is generally
explained by the presence of persistent structures in the carrier
flow. The latter are responsible for a deterministic motion of the
particles during which their inertia affects their relative motion
with respect to the carrier flow: phenomenology indicates that heavy
particles are ejected from eddies while light particles tend to
concentrate in their cores.  This is indeed what our local analysis
predicts.  However we shall see in the next section that, although
preferential concentrations of inertial particles are triggered by the
local mechanisms just discussed, it is the dissipative character of
the dynamics which is solely responsible for the eventual strong
clustering of particles.

\section{Threshold in Stokes number for the formation of fractal
  clusters}
\label{sec:threshold}

\subsection{Dissipative dynamics and Lyapunov exponents}
\label{subsec:diss-lyap}

The temporal evolution of the separation $\R(t)$ between two
infinitesimally close trajectories of the phase space can be expressed
in terms of the Green function ${\cal J}_t$ associated to the
linearised dynamics (\ref{eq:tangentsys}). We indeed have
\begin{equation}
  \R(t) = {\cal J}_t \,\, \R(0),\qquad {\cal J}_t \equiv
    {\cal T}\!\!\exp \int_0^t {\rm d}\tau\,{\cal M}_\tau,
  \label{eq:defjacobi}
\end{equation}
where ${\cal T}\!\!\exp$ denotes the time-ordered exponential of
matrices and ${\cal M}_t$ is the evolution matrix defined in
(\ref{eq:tangentsys}). The symmetrical matrix ${\cal
J}_t^\intercal{\cal J}_t$ has positive eigenvalues that can be written
$\e^{2 \mu_1(t) \,t}, \dots, \e^{2 \mu_{2d}(t) \,t}$. The $\mu_j$'s
are called the \emph{stretching rates} (or the finite-time Lyapunov
exponents) and are generally labelled in non-increasing order
$\mu_1(t)\ge\cdots\ge\mu_{2d}(t)$. They measure the time-evolution of
infinitesimal volumes of phase space.  $\mu_1$ measures the
exponential growth rate of the distance between two neighbouring
trajectories, $\mu_1+\mu_2$ measures that of areas defined by three
trajectories, \etc The sum of the $2d$ stretching rates controls the
time evolution of phase-space $2d$-dimensional volumes and is
expressible in terms of the trace of the Green matrix ${\cal J}_t$.
It is easily shown that $\mu_1+\cdots+\mu_{2d}=-d/\Stok$.

The long-time behaviour of the local dynamics is dominated by the
almost-sure convergence of the stretching rates to the classical
\emph{Lyapunov exponents} $\lambda_j = \lim_{t\to\infty}\mu_j(t)$.
The multiplicative ergodic theorem \cite[]{o68} ensures that, under
some ergodicity hypothesis on the dynamics, the Lyapunov exponents are
independent of both the realisation of the random carrier flow and of
the peculiar trajectory $\X(t)$ around which the linearised
dynamics is considered.  The Lyapunov exponents are linked to many
fundamental features of the dynamics.  For instance, when the largest
Lyapunov exponent $\lambda_1$ is negative, the stability of the
linearised system is ensured and all the particle trajectories are
converging together, thereby leading to a somewhat degenerate
statistical steady state in which all the mass is concentrated in
discrete time-dependent points in phase space. However when
$\lambda_1$ is positive, the situation is more complex: the dynamics
is then said to be chaotic and stability is ensured only if the
initial phase-space separation $\R(0)$ is orthogonal to the
subspace generated by the eigendirections associated to the positive
Lyapunov exponents.

When $\lambda_1>0$ the long-time dynamics has richer features, characterised by the
convergence of the particle trajectories to complex dynamical
structures called \emph{attractors}. These are generally (evolving) fractal
sets of the phase space and can be characterised by their Hausdorff
dimension $d_H$, which is expected to coincide with the box-counting
dimension in non-degenerate cases. The appearance of attractors can be
seen as the result of a competition between stretching and folding
effects that occur during the chaotic motion of the particles. As a
consequence, the properties of these attractors, and in particular
their dimension, depend on the stretching rates of the dynamics. The
positive $\mu_j$'s are responsible for  stretching in their associated
eigendirections, while the negative rates give  contraction and hence
lead to folding.  Following this kind of approach, \cite{ky79}
proposed to estimate the dimension of the attractor by the Lyapunov
dimension, defined as
\begin{equation}
  d_L \equiv J - \frac{\lambda_1+\cdots+\lambda_J}{\lambda_{J+1}},
  \quad \mbox{where}\quad \lambda_1+\cdots+\lambda_J\ge0
  \quad\mbox{and}\quad \lambda_1+\cdots+\lambda_{J+1}<0.
  \label{eq:deflyapdim}
\end{equation}
This non-random number can be interpreted heuristically as the
dimension of phase-space objects that keep a constant volume during
time evolution. As already stated above, the long-time evolution of
$k$-dimensional volumes is governed by the sum of the $k$ first
Lyapunov exponents.
\begin{figure}
\centerline{\includegraphics[width=0.4\textwidth]{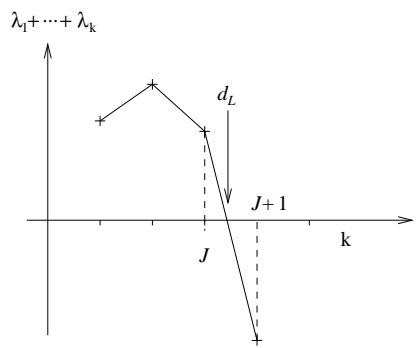}}
\caption{\label{fig:lyapdimdef} The Lyapunov dimension
  $d_L$ is defined by interpolating to non-integer values the
  exponential growth rate of phase-space $k$-dimensional volumes
  obtained from the sum of the $k$ largest Lyapunov exponents.}
\end{figure}
The Lyapunov dimension is obtained by interpolating this sum to
non-integer values and determining where it vanishes (see
figure~\ref{fig:lyapdimdef}).  It was shown by \cite{do80} that it
serves as a rigorous upper-bound for the Hausdorff dimension $d_H$;
this will be used in next subsection to show that below a critical
value of the Stokes number, the particles form fractal clusters in the
position space. As we shall see in \S\ref{sec:mass}, the Lyapunov
dimension is actually equal to the information dimension associated to
the steady-state phase-space density of particles and is thus related
to the small-scale properties of the mass distribution of inertial
particles.

\subsection{A heuristic approach in two dimensions}
\label{subsec:heuristics}

We now give a heuristic argument which predicts the presence of a
threshold in Stokes number for clustering of heavy particles.  This
result, already described briefly in \cite{b03}, is here presented
in  more details.

As already stated, the particles concentrate on dynamical attractors
in the position--covelocity phase space. The main idea of the argument
presented here comes from a trivial observation.  Focusing on
positions while ignoring covelocities amounts to projecting the
attractor from the $2d$-dimen\-sional phase space onto the
$d$-dimensional space of particle positions.  A standard result on the
geometry of fractal sets \cite[see, e.g.,][]{f86} is that, when the
Hausdorff dimension $d_H$ of the attractor is less than the dimension
of the projection space (here, $d$), the projection of the fractal is
itself a fractal set with Hausdorff dimension $d_H$. If however
$d_H>d$, the projection has dimension $d$.  Now, using the Lyapunov
dimension $d_L$ as a rigorous upper bound of the fractal dimension
$d_H$ of the phase-space attractor, the condition $d_L<d$ is a
sufficient condition for particles to form fractal clusters in
position space.  The next step is to find a condition ensuring that
the Lyapunov dimension is less than the space dimension. From its
definition (\ref{eq:deflyapdim}), it is clear that $d_L<d$ if the sum
of the $d$ largest Lyapunov exponents is negative.  Hence, the problem
reduces to finding a sufficient condition for
$\lambda_1+\cdots+\lambda_d<0$.

To estimate  the $d$ largest Lyapunov exponents we first
consider the \emph{stability exponents} defined as
\begin{equation}
  \mu_j \equiv \lim_{t\to\infty} \frac{1}{t} \ln \alpha_j(t),
  \label{eq:defstability}
\end{equation}
where the $\alpha_j(t)$'s are the eigenvalues of the Jacobi matrix ${\cal J}_t$ 
defined in
(\ref{eq:defjacobi}), which are usually complex
numbers.  Although there is no equivalent of the Oseledets
ergodic theorem for such complex  exponents, it is usually supposed that the
limit in (\ref{eq:defstability}) exists in generic situations and
depends neither on the realisation of the carrier velocity field, nor
on the particular trajectory along which they are calculated \cite[see,
e.g.][]{gso87}.  The real parts of the stability exponents generally
differ from the Lyapunov exponents.  It is however possible to obtain
inequalities between these two sets of exponents, using Browne's
theorem \cite[see, e.g., for details ][]{mr92}.  We will also need the
strain exponents $\eta_\ell$ analogous to the $\mu_j$'s but this time
associated to the $(d\times d)$-dimensional matrix ${\cal T}\!\!\exp
\int_0^t {\rm d}\tau\,\bm \sigma(\tau)$.

We now focus on the case of heavy particles ($\beta<1$) embedded in a
two-dimensional incompressible carrier flow.  Note first that, as we
have seen from the local analysis in the previous section, the heavy
particles tend to concentrate in the filamentous structures of the
flow (hyperbolic regions).  The eigenvalues associated to the local
dynamics of the particles are given by the quadratic relation
(\ref{eq:2ndordervp}) in terms of those associated to the strain
matrix of the carrier flow.  Now comes the heuristic point. 
Let us assume that the same quadratic
relation allows us to express, at least in an approximate way, the
stability exponents $\alpha_j$'s as a function of the strain exponents
$\eta_\ell$'s defined above\footnote{This assumption would be exact if
one could replace time-ordered exponentials appearing in the
definition of the various Green functions by ordinary exponentials; of
course, this is not the case here.}.  The latter are expected to be
real since they are dominated by the hyperbolic regions of the flow.
In two dimensions there are only two strain exponents $+\eta$ and
$-\eta$.  Under these hypotheses a sufficient condition for fractal
clustering of the particles can be written
\begin{equation}
  \Stok \le \frac{1}{\beta^2\,\eta} \,\left(
  \beta-2+2\sqrt{1-\beta+\beta^2} \right).
  \label{eq:critical1}
\end{equation}
Observe that the (non-dimensional) strain exponent $\eta$ is calculated
along the path of inertial particles and not along that of fluid
particles.  As the heavy particles spend a long time in the hyperbolic
regions, the strain should satisfy $\eta \ge \lambda_f\,L/U$, where
$\lambda_f$ denotes the positive Lyapunov exponents associated to the
dynamics of fluid particles.  Using this relation in
(\ref{eq:critical1}) we obtain that there is clustering if
\begin{equation}
  \Stok \le \frac{U}{\beta^2\,\lambda_f\,L} \,\left(
  \beta-2+2\sqrt{1-\beta+\beta^2} \right).
  \label{eq:critical2}
\end{equation}
Note that this sufficient condition gives heuristic evidence that
clustering already occurs at small Stokes numbers, although the Stokes
drag is then very strong, and thus the inertial particle motion is very
close to that of simple passive tracers.  This is consistent with
predictions made by \cite{bff01}.  Their approach is based on the
observation of \cite{m87} that the discrepancy between inertial
particle dynamics and passive tracers occuring when $\Stok\ll1$ can be
captured by a spatial Taylor expansion of the particle velocity field,
leading to a model in which the particle is advected by a synthetic
flow comprising a small compressible component.  Such an expansion
allowed Balkovsky \etal \ to show that moments of particle density
grow exponentially in times, which is indeed indicative of the formation of
clusters.

The existence of a threshold in Stokes number for particle clustering
requires that for large-enough values there be no clustering.  It is
easily checked from equations (\ref{eq:newton}) governing
inertial-particle dynamics that, when increasing the Stokes number,
the particle are less and less influenced by the carrier flow.  More
precisely, in the limit $\Stok\to\infty$, their motion becomes purely
ballistic and their distribution in position space tends to become
uniform, ensuring absence of fractal clusters for large-enough Stokes
number.

Note that the same type of arguments when applied to light particles
(with $\beta>1$) whose dynamics is dominated by elliptic regions of
the flow, would predict that they always concentrate on points located
in the cores of vortices (clusters of dimension $d_H=0$). As we shall
see from the numerical experiments of the next subsection, there are
ranges of values of the parameters $\beta$ and $\Stok$ for which this
strong clustering occurs in two dimensions, but for other values the
light particles can also form either fractal clusters or fill the
whole space.  This indicates the limits of our heuristic approach. The
arguments used in the case of heavy particles do not apply to light
particles in the elliptic regions since the eigenvalues of the local
dynamics are then complex and cannot be used to estimate the stability
exponents. Moreover, in order to match the limit $\Stok\to0$ in which
the motion of light particles recovers that of fluid particles, it is
clear that their dynamics is influenced by their stay in the
hyperbolic regions, even if this stay is shortened by inertial
effects.  The phenomenological derivation of the sufficient
condition~(\ref{eq:critical2}) was just meant to give an idea of the
physical mechanisms responsible for particles clustering. Its
qualitative validity is however confirmed numerically, as we shall now
see.
 
\subsection{Numerical evidence for fractal clustering in two and three
  dimensions}
\label{subsec:numerics}

In our simulations the inertial particles positions are confined to a
periodic domain of size~$L$. The prescribed carrier flow is a solution
to the forced Stokes equation
\begin{equation}
  \partial_t \bm u = \nu \Delta \bm u + \bm f(\bm x,t), \qquad
  \nabla\cdot\bm u = 0,
  \label{eq:Stokes}
\end{equation}
where $\bm f(\bm x,t)$ is a space-periodic, isotropic and homogeneous
Gaussian random forcing concentrated at spatial scales of the order of
the box size $L$ and delta-correlated in time. Such a carrier flow
provides control over the statistical properties of the Gaussian velocity
field $\bm u$ and allows accurate numerics.
In practice we take   $\nu=1$ and assume that the velocity field has
the lowest reasonable number of spatial Fourier modes for which
isotropy is approximately ensured: 9 in two dimensions and 27 in three
dimensions\footnote{Fewer Fourier modes may actually lead to
degenerate situations and in particular to insufficiently mixing
dynamics.}. Note that two-dimensional flows satisfying
(\ref{eq:Stokes}) were also considered by \cite{ss02} who proved in
this setting existence of the random dynamical attractor for very
heavy particles.

The individual trajectories of inertial particles are integrated using
a fourth-order Runge--Kutta scheme. Choosing a carrier fluid velocity
with rather few Fourier modes makes it unnecessary to interpolate the
velocity at particle locations, since we can calculate it by direct
summation of the Fourier series. Fully resolving the finest structures
of the spatial particle distribution requires a strict control over
the small-scale dynamics, as is here the case.

We present in this section measurements of the Lyapunov exponents
associated to the dynamics of the particles.  To estimate these
quantities, we use the standard method developed by \cite{bggs80} -
see also the book by \cite{cpv93} for a precise description of this
method. The idea is to integrate the time evolution of $2d$
infinitesimal separations $\bm R_1,\dots,\bm R_{2d}$ governed by the
linearised dynamics (\ref{eq:tangentsys}) and along the path of a
given particle.  The Lyapunov exponents are obtained from the
exponential growth rates of the distances, surfaces, volumes, \etc
defined by these infinitesimal vectors.  In order to prevent numerical
errors from accumulating and the length of the $\bm R_j$'s from
increasing too rapidly, their time evolution must be accompanied by
frequent renormalisation,\footnote{Here the term ``renormalisation''
is to be taken literally, meaning just ``making the norm finite''.}
using, for instance, the Gram--Schmidt procedure.  In practice, to
obtain the Lyapunov exponents for given values of the parameters
$\beta$ and $\Stok$, we integrate the linearised system for $10^5$
turn-over times performing a renormalisation of the vectors $\bm
R_1,\dots,\bm R_{2d}$ every tenth of turn-over time.

\begin{figure}
  \centerline{\subfigure[\label{fig:diag1}$d=2$]
    {\includegraphics[width=0.45\textwidth]{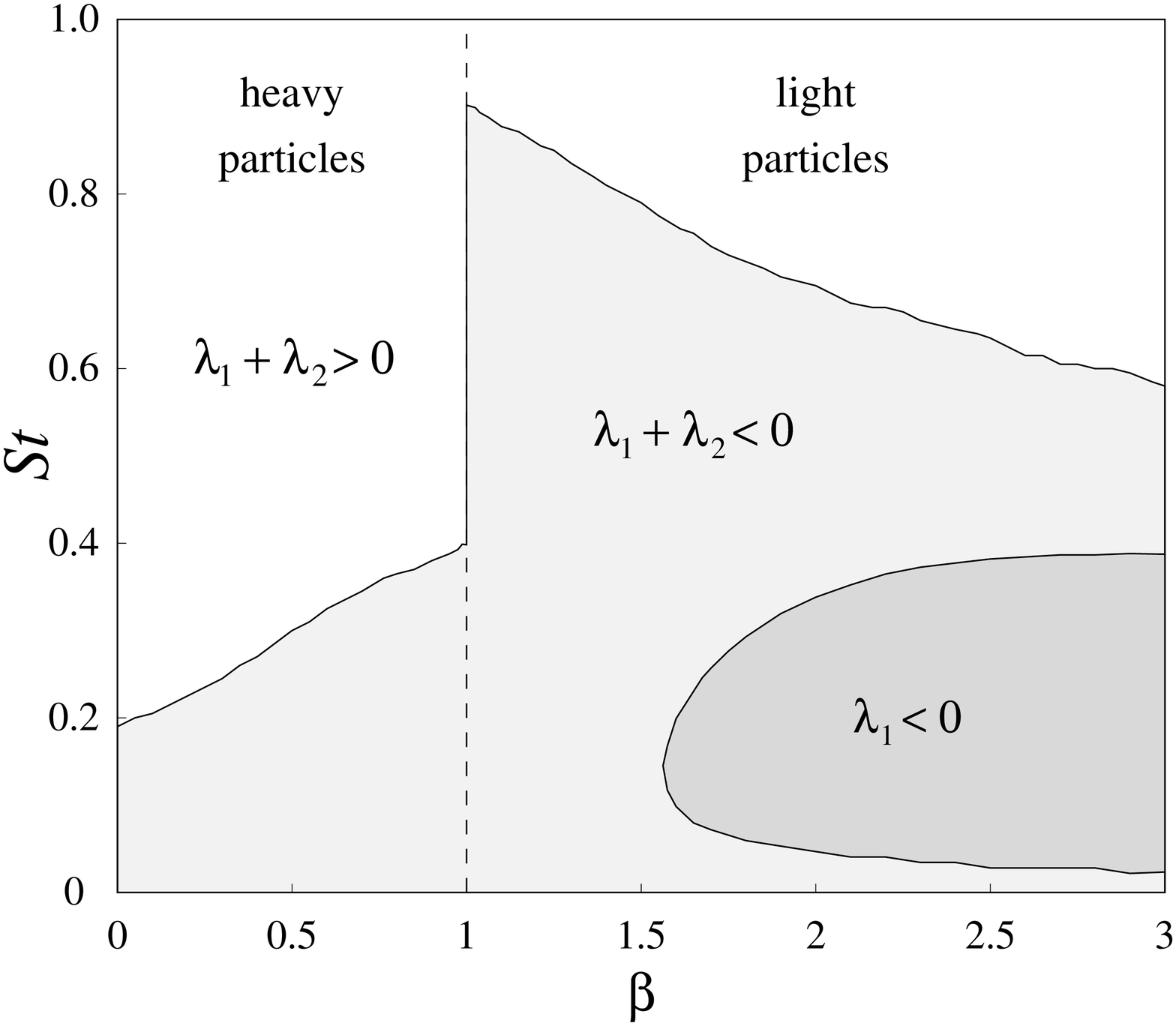}} \qquad
    \subfigure[\label{fig:diag2}$d=3$]
              {\includegraphics[width=0.45\textwidth]{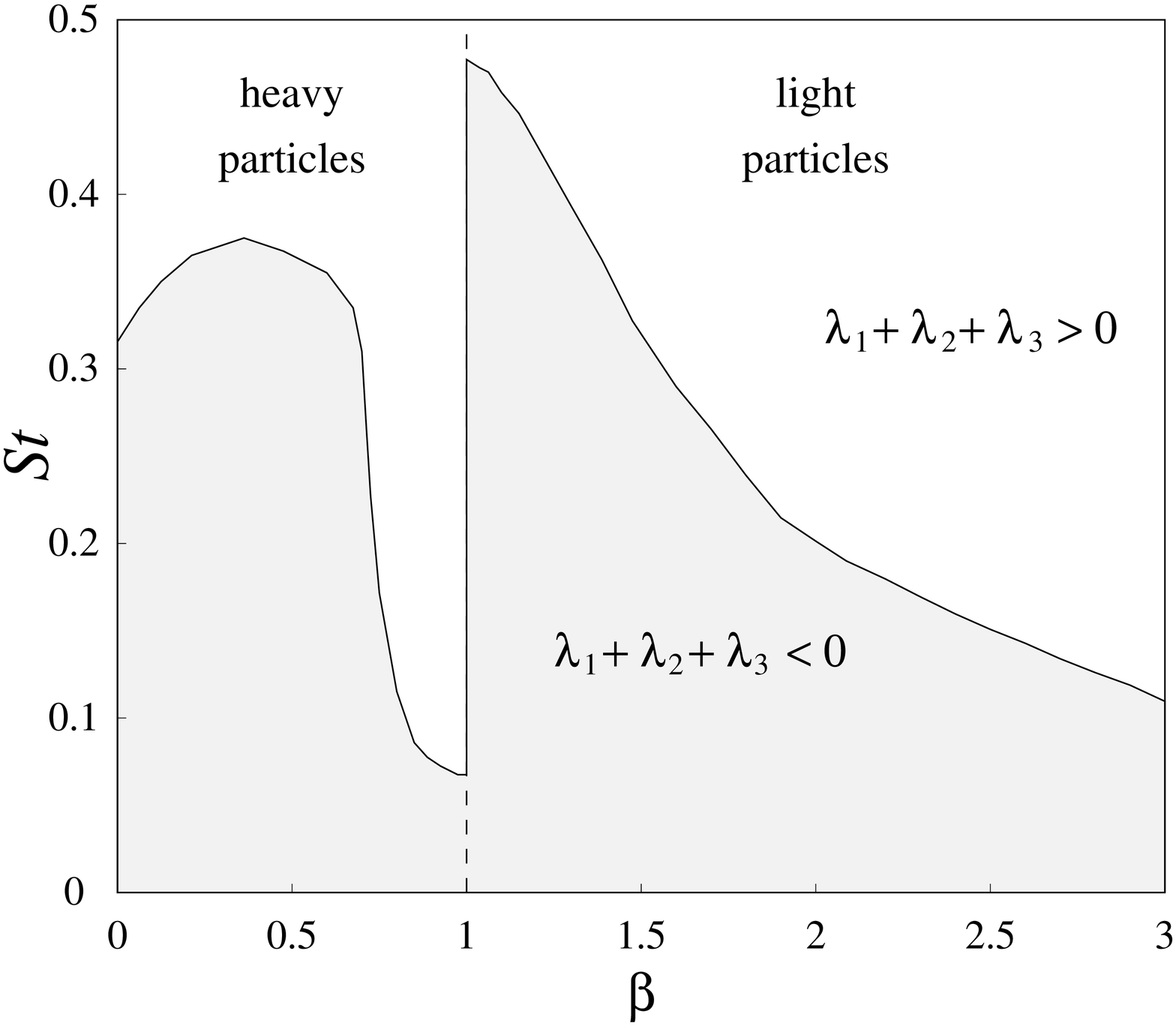}}}
\caption{\label{fig:diagramme} Phase diagrams in the parameter space
  $(\beta, \Stok)$ for $d=2$ (\textit{a}) and $d=3$ (\textit{b})
  representing the different regimes of the dynamics corresponding to
  different behaviours of the Lyapunov exponents.  For parameter
  values in the white area (such that $\lambda_1+\cdots+\lambda_d>0$),
  the distribution of particle fills the whole position space. In the
  light-gray area the Lyapunov dimension is less than $d$ (i.e.\
  $\lambda_1+\cdots+\lambda_d<0$) and particles concentrate on fractal
  clusters.  The darker gray area which is only present in the
  two-dimensional case corresponds to the regime for which all
  Lyapunov exponents are negative: the asymptotic distribution of
  particles is then atomic.}
\end{figure}
Figure \ref{fig:diagramme} (\textit{a}) shows the phase diagram
obtained numerically in the \emph{two-dimensional case}, which divides
the parameter space $(\beta, \Stok)$ into three different regimes.
The light-gray area represents those values for which the sum of the
two first Lyapunov exponents is negative.  The Lyapunov dimension is
there smaller than the dimension $d=2$ of the position space and the
particles form fractal clusters.  The second regime appearing in the
dynamics is when the sum of the two first Lyapunov exponents is
positive (white area).  This corresponds to the case when the Lyapunov
dimension is larger than two and the particles fill the whole domain.
The last regime, represented by the dark-gray area in figure
\ref{fig:diagramme} (\textit{a}), corresponds to values of the
dynamical parameters for which the largest Lyapunov exponent
$\lambda_1$ is negative. This strong clustering happens only when the
particles are lighter than the fluid.  They form there point-like
clusters and the associated mass distribution is said to be
\emph{atomic}.  In the case of bounded domains (such as considered
here) it corresponds to the large-time convergence of all particle
trajectories towards a single trajectory.

In \emph{three dimensions}, the situation is rather different. As
shown in figure~\ref{fig:diagramme} (\textit{b}), there are only two
different possible regimes in the particle dynamics: one corresponding
to a Lyapunov dimension larger then the position-space dimension and
in which particles fill the whole space; the other one associated to a
Lyapunov dimension smaller than $d$ and where particles concentrate
onto fractal clusters.  Surprisingly, there is no evidence for the
presence of a region in the parameter plane where the particles
cluster on a set of dimension less than two.  A systematic
investigation for a large number of different values of the parameters
$\beta$ and $\Stok$ was done. In particular, the Lyapunov exponents
were calculated for forty different values of $\Stok$ at $\beta=2.9$
for which clustering would have been expected.  Results indicate that
the sum of the two largest Lyapunov exponents always remains positive.
This observation contradicts the phenomenological understanding of
clustering which predicts that light particles may form clusters of
dimension between one and two in the core of the vortices.  The
absence of such a behaviour is not due to the choice of the model,
otherwise point-like clusters would also be absent in two dimensions.
A possible explanation is related to the properties of the carrier
flow considered here.  It could be that in our setting, the velocity
field $\bm u$ does not present rotational structures which are strong
enough and sufficiently persistent to permit such a clustering of the
particles. On the one hand, we have seen from the local analysis of
\S\ref{sec:local} that the elliptic regions of the carrier flow are
stable for light particles only if the Stokes number is sufficiently
small.  On the other hand, $\Stok$ has to be sufficiently large for
the particles to have enough inertia to separate from the fluid
motion. It may happen that these two conditions are here incompatible.
It would be of interest to confirm the absence of point-like clusters
by systematic investigations in different flow fields (and in
particular realistic ones); this will be done in future work.

The separatrix between the regimes associated to different signs of
the sum of the $d$ largest Lyapunov exponents has a discontinuity at
$\beta=1$ in both two and three dimensions.  This singular behaviour
is due to the degenerate dynamics associated to particles with the
same density as the fluid (such particles are usually referred to as
\emph{neutrally buoyant} particles).  It is indeed clear from
(\ref{eq:newton}) that when $\beta=1$, the covelocity does not depend
anymore on the velocity of the carrier flow; it simply relaxes
exponentially from its initial value to zero.  As a consequence, the
subspace of covelocities is an eigenspace for the linearised dynamics
associated to a Lyapunov exponent equal to $-1/\Stok$ and with
multiplicity $d$. Incompressibility of the carrier flow then implies
that $\lambda_1+\cdots+\lambda_d = 0$ on the whole line $\beta = 1$ of
the parameter space, which partly explains the singular behaviour
observed here.

\begin{figure}
\centerline{\includegraphics[width=0.55\textwidth]{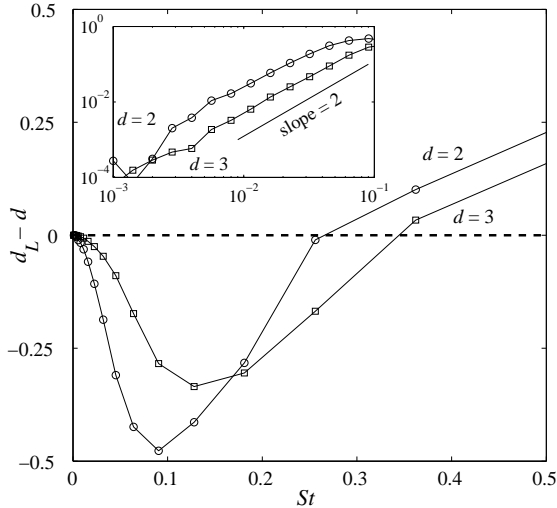}}
\caption{\label{fig:dimension} Difference $d_L-d$
  between the Lyapunov dimension associated to the dynamics of very
  heavy particles ($\beta=0$) and the position-space dimension, as a
  function of the Stokes number $\Stok$ (circles: $d=2$, squares:
  $d=3$). The critical Stokes number corresponds to the value for
  which $d_L=d$. Upper-left inset\,: same in log-log coordinates
  showing evidence for a quadratic behaviour at low Stokes numbers.}
\end{figure}
More quantitative information is provided by the behaviour of the
Lyapunov dimension $d_L$ as a function of the Stokes number. Figure
\ref{fig:dimension} shows the results obtained from both two- and
three-dimensional simulations in the case of very heavy particles
($\beta=0$). First, we observe that the dimension of the attractor
indeed tends to the space dimension $d$ at low Stokes numbers. More
precisely, in both two and three dimensions, the Lyapunov dimension
behaves in this asymptotic regime as $d_L \simeq d - C\,\Stok^2$.
Such a quadratic behaviour near vanishing Stokes numbers has been
obtained theoretically by \cite{bff01} using the method of advection
by a synthetic compressible flow, already mentioned.  At large Stokes
numbers, the Lyapunov dimension becomes larger and larger, eventually
reaching $2d$ when $\Stok\to\infty$.  Between these two asymptotic
regimes, there is a range of Stokes numbers for which the Lyapunov
dimension - and thus the fractal dimension of the attractor - is
smaller than $d$.  For such Stokes numbers, the particles form fractal
clusters. The graphs of the Lyapunov dimension against the Stokes
number, shown in figure~\ref{fig:dimension}, display a minimum around
$\Stok =0.09$ for $d=2$ and $\Stok = 0.14$ for $d=3$. These are the
Stokes numbers for which the strongest clustering takes
place. Finally, these experiments also confirm the existence of a
critical value for the Stokes number below which fractal clustering
occurs; on the figure it corresponds to the value for which the curves
cross the line $d=d_L$ (around $\Stok =0.24$ for $d=2$ and
$\Stok=0.32$ for $d=3$).

\begin{figure}
  \centerline{\subfigure[\label{fig:clust1}$\Stok=10^{-3}$]
    {\includegraphics[width=0.46\textwidth]{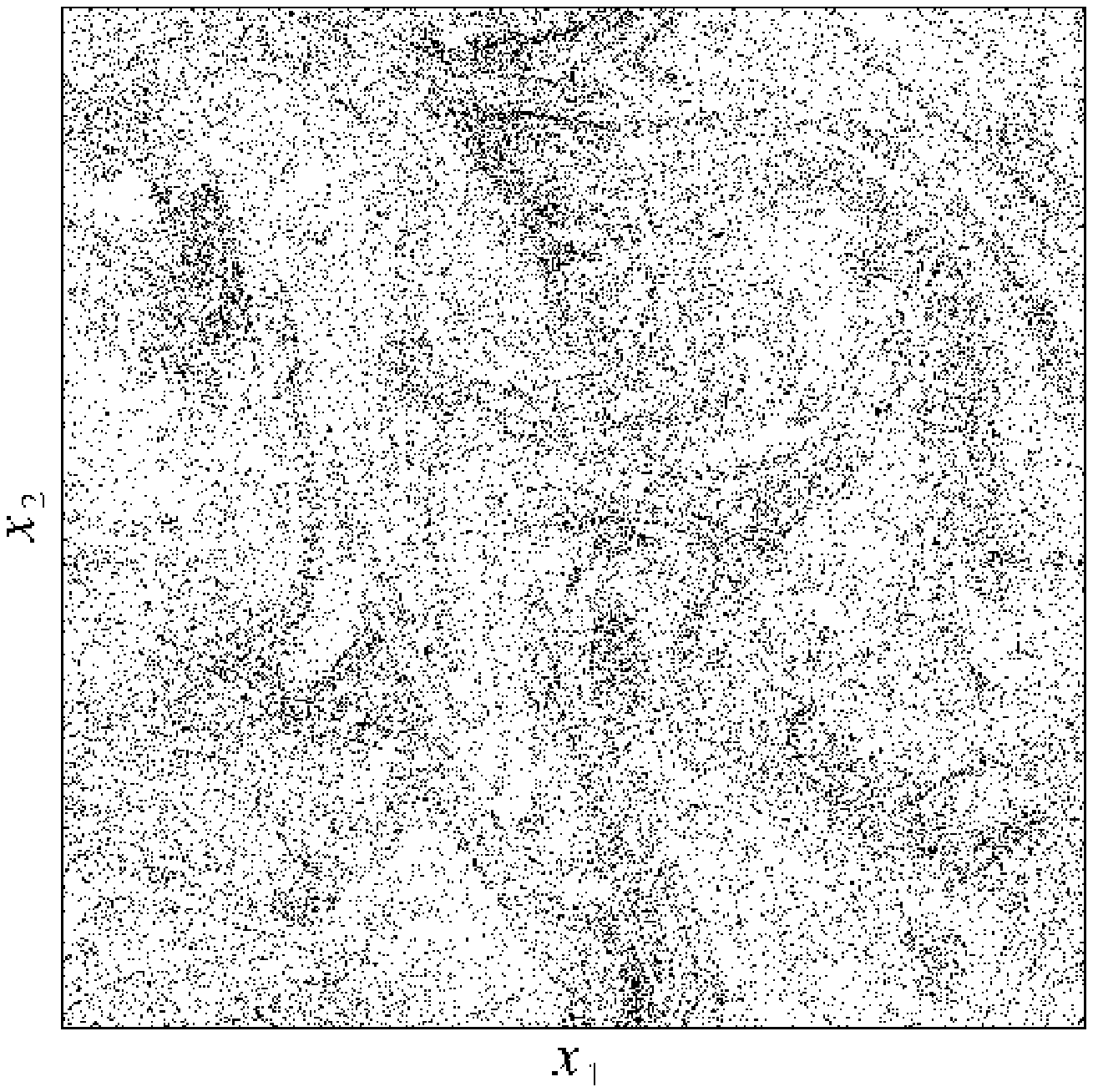}}
    \qquad \subfigure[\label{fig:clust2}$\Stok=10^{-2}$]
    {\includegraphics[width=0.46\textwidth]{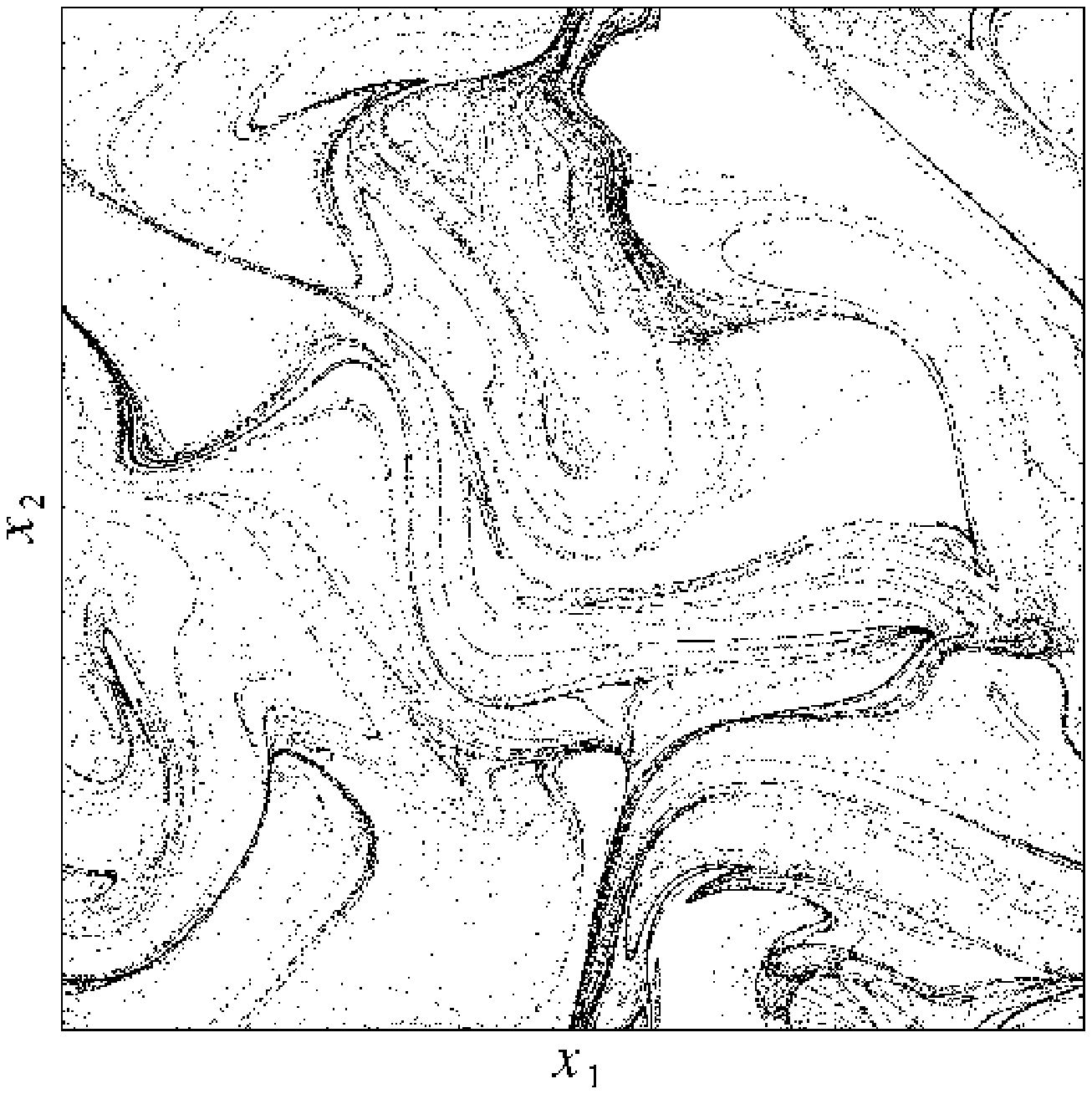}}}
    \centerline{\subfigure[\label{fig:clust3}$\Stok=10^{-1}$]
    {\includegraphics[width=0.46\textwidth]{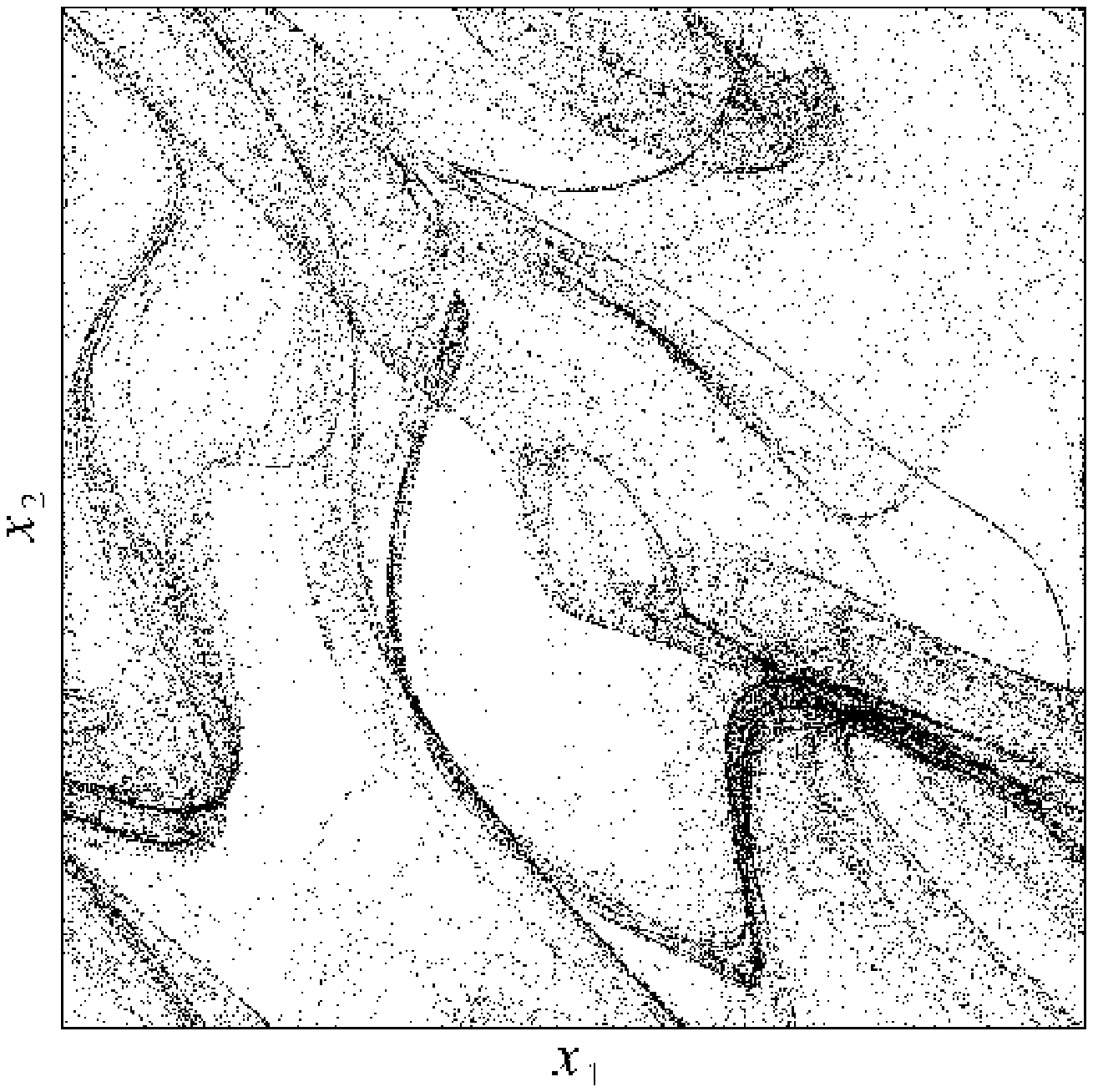}} \qquad
    \subfigure[\label{fig:clust4}$\Stok=1$]
    {\includegraphics[width=0.46\textwidth]{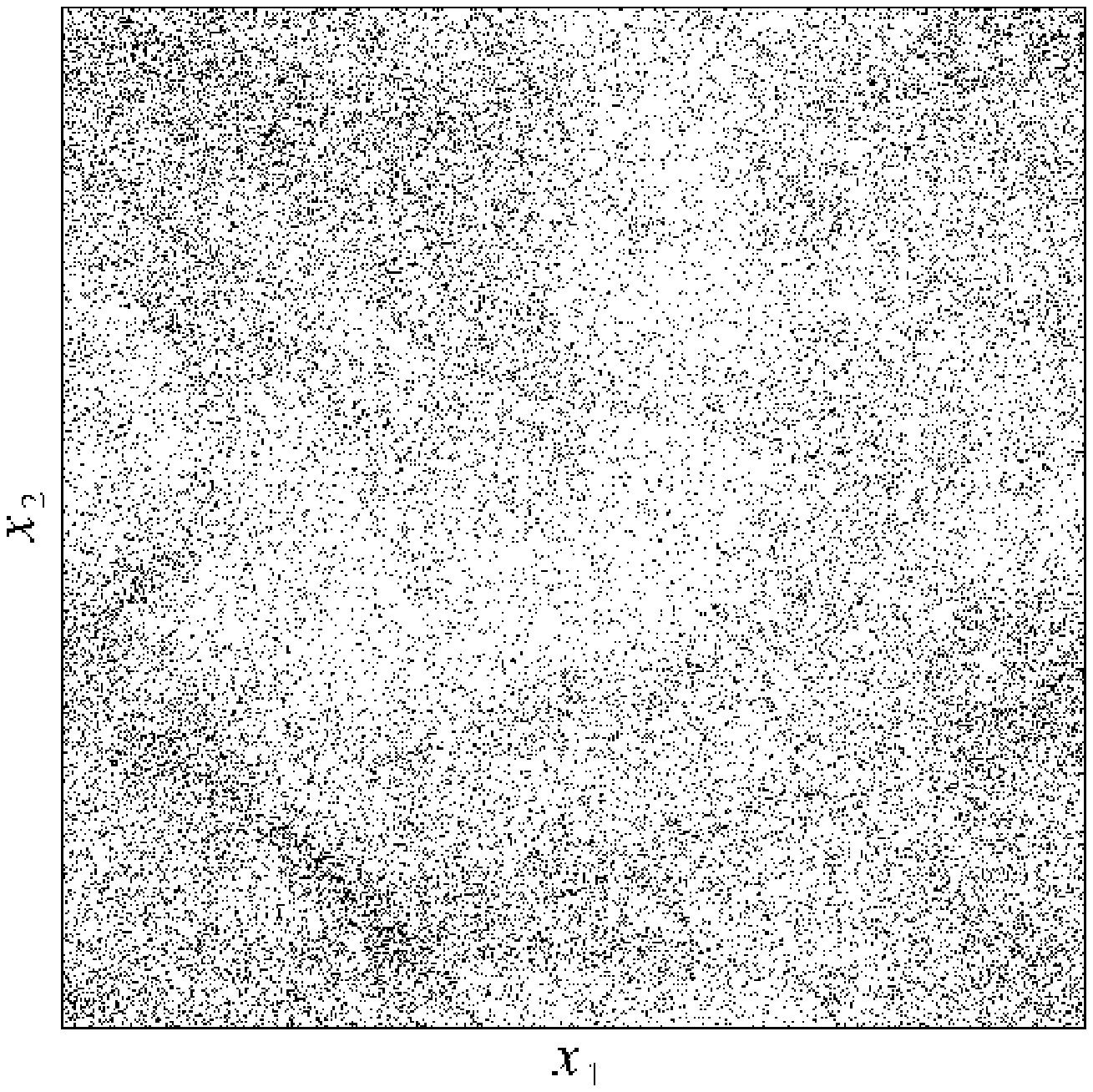}}}
  \caption{\label{fig:clusters} Snapshots of the positions of $N=10^5$
    heavy particles ($\beta=0$) associated to four different Stokes
    numbers as labelled. (\textit{a}), (\textit{b}) and (\textit{c})
    correspond to values smaller than the threshold, so that particles
    form fractal clusters. (\textit{d}) corresponds to a Stokes number
    larger than the critical value, so that the particles fill the
    whole domain.}
\end{figure}
Qualitative insight into clustering is provided by
figure~\ref{fig:clusters} where snapshots of the position of $10^4$
very heavy particles illustrate the different regimes arising in the
two-dimensional dynamics when increasing the Stokes number.  First,
when the Stokes number is very small (here $\Stok = 10^{-3}$ for
figure~\ref{fig:clusters} \textit{a}), the particles are rather
densely distributed but it is possible to observe fractal clusters
with a dimension very close to the space dimension $d$.  This picture
illustrates the highly singular nature of the vanishing Stokes number
limit. Note also that, as the Stokes number becomes very small, the
dynamical fractal clusters are becoming denser and more
space-filling. Conversely, as we increase $\Stok$, the dimension of
the clusters decreases and large empty areas appear
(figure~\ref{fig:clusters} \textit{b}). This process continues up to
the value of $\Stok$ corresponding to the strongest clustering. Above
this value, the dimension of clusters increases as a function of
$\Stok$ and caustic-like structures become conspicuous in the particle
spatial distribution (figure~\ref{fig:clusters} \textit{c}).  Finally,
when the Stokes number is larger than the critical value, the
particles fill the whole domain, but in a nonuniform way
(figure~\ref{fig:clusters} \textit{d}).

\section{Statistical properties of the mass distribution}
\label{sec:mass}

\subsection{Phase-space density fluctuations and anomalous scaling}
\label{subsec:anomalous}

In the previous section the Lyapunov dimension was used as an estimate
to determine the presence or not of fractal clusters of particles.
This dimension is not only an upper bound on the fractal dimension of
the dynamical attractor; it also provides additional information on
the small-scale properties of the particle distribution. Consider an
initial density of particles in the phase space which is uniform when
projected on the position space and may or may not possess some
dispersion in covelocities at each position.  At a given time $t$
sufficiently large to be close to the asymptotic regime, the
phase-space density field $f(\bm x, \bm v, t)$ is generally singular:
its support is exactly the random dynamical attractor toward which all
individual trajectories converge. This limiting density is a random
equivalent of the Sinai--Ruelle--Bowen (SRB) measure, well-known in
the study of dissipative dynamical systems \cite[see,
e.g.,][]{er85,y02}.  The density distribution along the attractor can
be characterised by the mass of particles contained in small balls of
radius $r$ centred on the attractor.  The requirement to be on the
attractor is automatically satisfied if we let the ball centre follow
a given trajectory $(\X(t), \V(t))$ in phase space.  Let us
thus introduce the phase-space mass in a phase-space ball or radius
$r$ around the trajectory:
\begin{equation}
  m_r (\X(t), \V(t), t) \equiv \int_{|\bm x|^2 + |\bm v|^2 \le r^2}
  {\rm d}\bm x \,{\rm d}\bm v\, f(\X(t)+\bm x, \V(t)+\bm v, t).
  \label{eq:defmassphase}
\end{equation}
Note that ${\bm x}$ and ${\bm v}$ are here non-dimensional variables,
so that it is meaningful to use the Euclidean norm in the $({\bm x},
{\bm v)}$ phase space.  Under suitable non-degeneracy hypotheses for
the random flow defined by the dynamics, it was shown by \cite{ly88}
that at small scales, $m_r$ behaves as $r^{d_L}$.  More precisely, for
almost every particle trajectory,
\begin{equation}
 \frac{\ln m_r (\X(t), \V(t), t)}{\ln r} \to d_L \quad
 \mbox{as}\quad r \to 0.
  \label{eq:ledryoung}
\end{equation}
If the limit exists then it is equal to the information dimension of
the statistical-steady-state density $f$ (defined as the smallest
Hausdorff dimension of a set of non-vanishing mass).  In other words,
the information dimension is equal to the Lyapunov dimension, identity
which was conjectured by Kaplan \& Yorke and which was rigorously
proved by Ledrappier \& Young for SRB measures.  Of course, for a
finite radius $r$ of the ball, the mass of particles does not scale
exactly as $r^{d_L}$ but generally fluctuates.  These fluctuations can
be captured by investigating the scaling properties of the moments of
order $n$ of the mass, which follow power laws with exponents $\xi_n$
at small $r$'s:
\begin{equation}
 \left\langle\,m_r^n (\X(t), \V(t), t)\,\right\rangle \sim
 r^{\xi_n} \quad \mbox{when} \quad r \ll L. \label{eq:defxin}
\end{equation}
Here $\langle\cdot\rangle$ denotes averaging with respect to the
realisations of the carrier flow.  These exponents can be expressed in
terms of the spectrum of dimensions $D_n$ \cite[see][]{g83,hp83} by
$\xi_n = n\,D_{n+1}$.  The result of Ledrappier \& Young clearly
implies that $({\rm d}\xi_n/{\rm d}n)|_{n=0} = d_L$.  There is yet
another exact result which relates the exponents $\xi_n$'s to the
linearised dynamics. \cite{bs88} showed that the exponent $\xi_1$ is
given by the Lyapunov moments dimension associated to the generalised
Lyapunov exponents \cite[see][]{bppv85}. This means that it satisfies
\begin{equation}
  \lim_{t\to\infty} \frac{1}{t} \,\ln \left\langle |\bm R(t)|^{-\xi_1}
  \right\rangle = 0,
\end{equation}
where the vector $\bm R(t)$ denotes the phase-space separation between
two infinitesimally close trajectories.  Here we observe that
phenomenological arguments are proposed in \cite{bgh03} to relate the
full set of scaling exponents to the linearised system
(\ref{eq:tangentsys}) governing the time evolution of the
infinitesimal separation $\bm R(t)$.  More precisely, for an arbitrary
order $n$, the scaling exponent $\xi_n$ can be expressed in terms of
the entropy function governing the long-time fluctuations of the
stretching rates around their limiting values, the latter being
precisely the Lyapunov exponents.  We shall return to these matters
and their implications in the conclusion (\S\ref{sec:conclusion}).

Note that a self-similar distribution of particles would have $\xi_n =
n\,d_L$.  Because of the convexity of the function $n\mapsto \xi_n$ (a
consequence of a H\"older's inequality applied to the moments of
mass), it is clear that a necessary and sufficient condition for the
exponents to depend \emph{nonlinearly} on the order $n$ is that
$\xi_1<d_L$. Such a nonlinear behaviour is frequently referred to as
\emph{anomalous scaling} since it cannot be predicted by dimensional
analysis.  It implies that the particle mass distribution is
\emph{intermittent} and in particular that its probability density
function (PDF) does not have Gaussian tails. More specifically,
according to the usual multifractal formalism \cite[see,
e.g.,][]{f95}, we can write the stationary PDF of the phase-space mass
$m=m_r(\X(t), \V(t), t)$ approximately as
\begin{equation}
  p_r(m) = \frac{C}{m\,\ln r} \, \exp \left \{ (\ln r)\, \left[2d -
  D\left(r,\, \frac{\ln m}{\ln r}\right) \right]\right\}.
  \label{eq:defpr}
\end{equation}
The scaling (\ref{eq:defxin}) implies that in the limit $r\to0$ with
$(\ln m)/(\ln r) = h$ fixed, the function $D(r,h)$ tends to a limit that
we denote $D_0(h)$. The scaling exponents $\xi_n$ are related to
$D_0(h)$ by a Legendre transform
\begin{equation}
  \xi_n = 2d + \inf_h \left[\, h\,n - D_0(h) \,\right].
  \label{eq:legendre}
\end{equation}
The limit $D_0(h)$ is frequently referred to as the \emph{multifractal
  spectrum} of the random dynamical attractor. The names entropy
function, rate function or Cram\'{e}r function associated to the
mass fluctuations are also sometimes used to refer to
$2d-D_0(h)$. The multifractal spectrum can be interpreted
heuristically as the dimension of the set on which the mass in small
balls scales as $r^h$.  A straightforward consequence of the
almost-sure scaling of mass (\ref{eq:ledryoung}) is that the maximum 
of  $D_0(h)$ is $2d$ and  is attained for $h=d_L$.  Knowledge
of $D_0(h)$ gives  a very fine description of  the small-scale properties
of the mass distribution in phase space.

\subsection{Distribution of particle positions at low Stokes
  numbers}
\label{subsec:lrgdev}

In many applications and in particular those where interactions
between inertial particles are introduced (such as collisions or
chemical reactions), it is of greater interest to characterise not the
phase-space mass but the position-space distribution of particles. To obtain
the spatial density $\rho(\bm x,t)$ one must integrate the phase-space
density $f$ over the covelocities.  As seen in
\S\ref{subsec:heuristics}, when the dimension of the attractor is
smaller than $d$, the density $\rho$ is singular with support on the
projection of the attractor, that is on clusters.  To investigate the
distribution of particles, we consider a small ball of radius $r$
centred on a given position $\bm x$ which does not necessarily
correspond to the position of a particle and hence may not intersect
the attractor.  The mass of particles contained in this ball is
\begin{equation}
  \overline{m}_r (\bm x, t) \equiv \int_{|\bm y| \le \rho} {\rm d}\bm
  y \,\rho(\bm x + \bm y,t) = \int_{|\bm y| \le \rho} {\rm d}\bm y
  \int {\rm d}\bm v\, f (\bm x, \bm v, t).
  \label{eq:defmassphys}
\end{equation}
Similarly to the phase-space mass distribution, the moments of
$\overline{m}_r$ are expected to display anomalous scaling.  The usual
multifractal formalism used in the previous subsection leads to
defining the multifractal spectrum $\overline{D}_0(h)$ associated to
the fluctuations of the position-space mass $\overline{m}_r$.
Relating the distribution of $\overline{m}_r$ to that of the mass in
phase space $m_r$ requires an integration over covelocities. This can
generally not be done without a better understanding of the fractal
properties of the attractor in the covelocity directions. It is clear
that the position variables and the covelocity variables do not play
similar roles in the dynamics, so that we have a kind of anisotropy in
the phase-space dynamics.  This prevents us from using standard tools
of the geometry of fractal sets when integrating over the
covelocities.

\begin{figure}
\centerline{\includegraphics[width=0.4\textwidth]{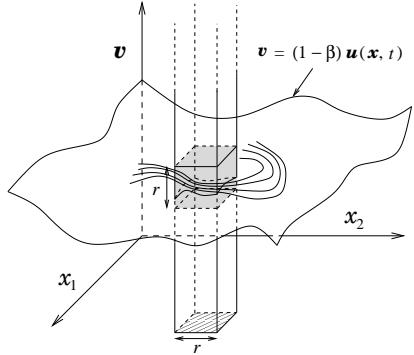}}
\caption{\label{fig:fractal} Sketch illustrating the
  arguments used at low Stokes number for the determination of the
  mass of particles $\overline{m}_r$ in a small volume of size $r$ in
  the position space. The case $d=2$ is here assumed and the two
  components of the covelocity in the four-dimensional phase space are
  schematically represented as a single variable. }
\end{figure}

This matter does however simplify if we limit ourselves to the
asymptotics $\Stok\ll1$.  The position-space mass distribution can
then be obtained at intermediate spatial scales from the mass
distribution in the full phase space.  Indeed, the covelocities of the
particles associated to low Stokes numbers are within a distance order
$\Stok$ of the phase-space manifold $\bm v = (1-\beta)\,\bm u(\bm
x,t)$. Hence, when $\Stok \ll r/L \ll 1$, the position-space mass
$\overline{m}_r$ is given by the particles contained in a phase-space
volume also of size $r$ centred on this manifold (see figure
\ref{fig:fractal}). We thus introduce the exponents $\zeta_n$ for this
intermediate asymptotic range:
\begin{equation}
  \left\langle\overline{m}_r\,^n\right\rangle \sim r^{\zeta_n} \quad
  \mbox{for} \quad \Stok \ll r/L \ll 1.
  \label{eq:defzetan}
\end{equation}
We can relate the $\zeta_n$'s to the exponents $\xi_n$ associated to
the phase-space mass $m_r$, taking into account that $m_r$ is
evaluated along the attractor while $\overline{m}_r$ is not.  We have
seen in \S\ref{sec:threshold} that for $\Stok\ll1$, the particle are
located on dynamical clusters whose fractal dimension is equal to that
of the phase-space attractor. The probability that the ball of size
$r$ intersects the fractal cluster of particles is hence $\propto
r^{d-d_H}$.  When it intersects the projection of the attractor, the
mass inside the ball obviously behaves as $r^{\xi_n}$. We thus have
\begin{equation}
  \zeta_n = d-d_H + \xi_n.
  \label{eq:relexponents}
\end{equation}
The relation (\ref{eq:ledryoung}) of Ledrappier \& Young between the
information dimension and the Lyapunov dimension  implies that
$({\rm d}\zeta_n/{\rm d}n)|_{n=0} = d_L$. Moreover, the fact that $\xi_0 = 0$,
together with relation (\ref{eq:relexponents}), implies that $\zeta_0
= d-d_H$. Conservation of the total mass of particles implies that the
average mass in a ball of the position space scales as its volume. We
hence have $\zeta_1 = d$ and thus $\xi_1 = d_H$.  Note that because of
the convexity of $n\mapsto\xi_n$, a necessary and sufficient condition
to have multifractal clusters of particles (that is a non-linear
dependence of the $\zeta_n$) is that $d_H<d_L$ and, of course,  
a value of the Stokes number below the critical clustering value. 

\cite{hk97} showed that the dimension spectrum $D_n$ of fractal
measures defined in the previous subsection is preserved under typical
projections when $1<n\le2$ (``typical projections'' means here almost
all of them).  From a direct application of this result to the
steady-state phase-space density $f$, one expects that for order-unity
values of the Stokes number, $\zeta_n = d-d_H + \min(n\,d,\xi_n)$ for
$0<n\le1$.  When $\Stok$ is below the critical value, it is clear from
convexity of the $\xi_n$'s together with $({\rm d}\xi_n/{\rm
d}n)|_{n=0} = d_L<d$ that $\xi_n < n\,d$ for all $n$. This implies
that the relation (\ref{eq:relexponents}) may be extended to finite
values of $\Stok$. However, it is not yet completely clear whether or
not this result is applicable in our setting.  Indeed, because of the
aforementioned anisotropy of the phase-space dynamics, it is
conceivable that the projection onto the position space is not a
typical one.

\begin{figure}
\centerline{\includegraphics[width=0.65\textwidth]{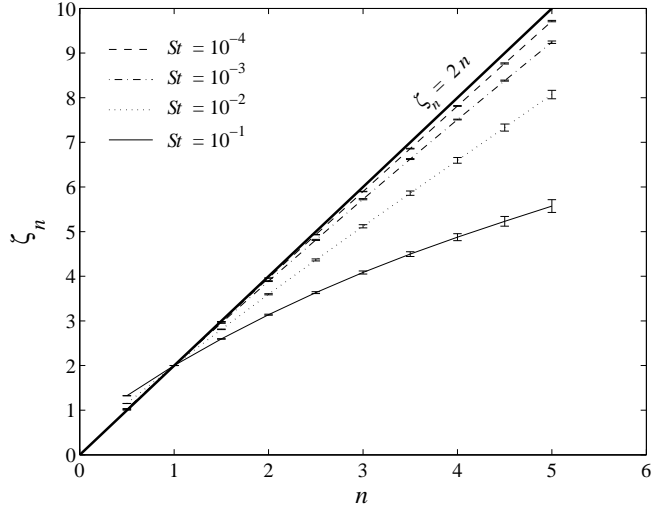}}
\caption{\label{fig:exposants} Scaling exponents $\zeta_n$ of the
  moments of the mass of particles in a small ball of radius $r$ as a
  function of their order $n$. The mass was calculated by box-counting
  with a total of $N=10^5$ particles and for four different values of
  the Stokes number, as labelled. The solid line represents the
  exponents associated to a uniform distribution of particles.}
\end{figure}
Our two-dimensional numerical experiments show that in the low Stokes
number regime, the distribution of inertial particle positions has
indeed multifractal properties.  For $\Stok\ll1$, particle positions
almost fill the whole space and, as seen previously, their Lyapunov
dimension behaves as $d_L \simeq d - C\,\Stok^2$. The scaling
exponents $\zeta_n$ of the moments of their mass distribution are
hence very close to $d\,n = 2\,n$, namely those given by simple
tracers dynamics.  Obtaining strong evidence for multifractal
clustering requires an accurate determination of the exponents
$\zeta_n$; this in turn requires performing long averages over the
realisations of the carrier flow. Figure \ref{fig:exposants} shows the
scaling exponents $\zeta_n$'s obtained after time averaging the mass
distribution for more than $10^4$ turnover times of the carrier flow.
Here we take very heavy particles ($\beta\ll1$) and plot the scaling
exponents $\zeta_n$ for different values of the Stokes number ranging
from $10^{-4}$ to $10^{-1}$.  Our results show that, already at very
low Stokes number, the discrepancy from a space-filling distribution
is increasingly noticeable when increasing the order $n$ of the
moment.

\begin{figure}
  \centerline{\subfigure[\label{fig:pdfm1}$\Stok = 10^{-3}$]
    {\includegraphics[width=0.45\textwidth]{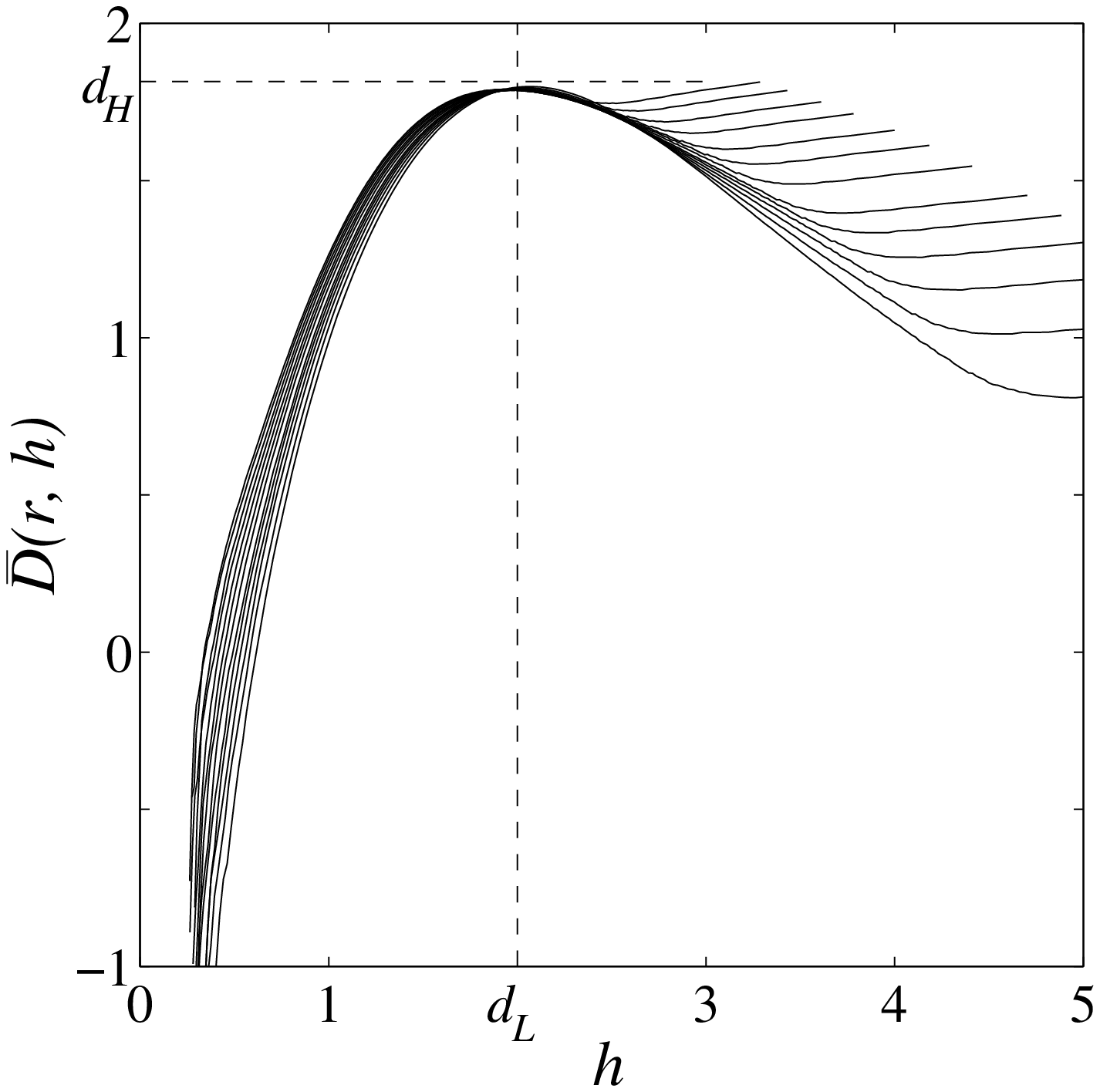}} \hfill
    \subfigure[\label{fig:pdfm2}$\Stok = 10^{-2}$]
              {\includegraphics[width=0.45\textwidth]{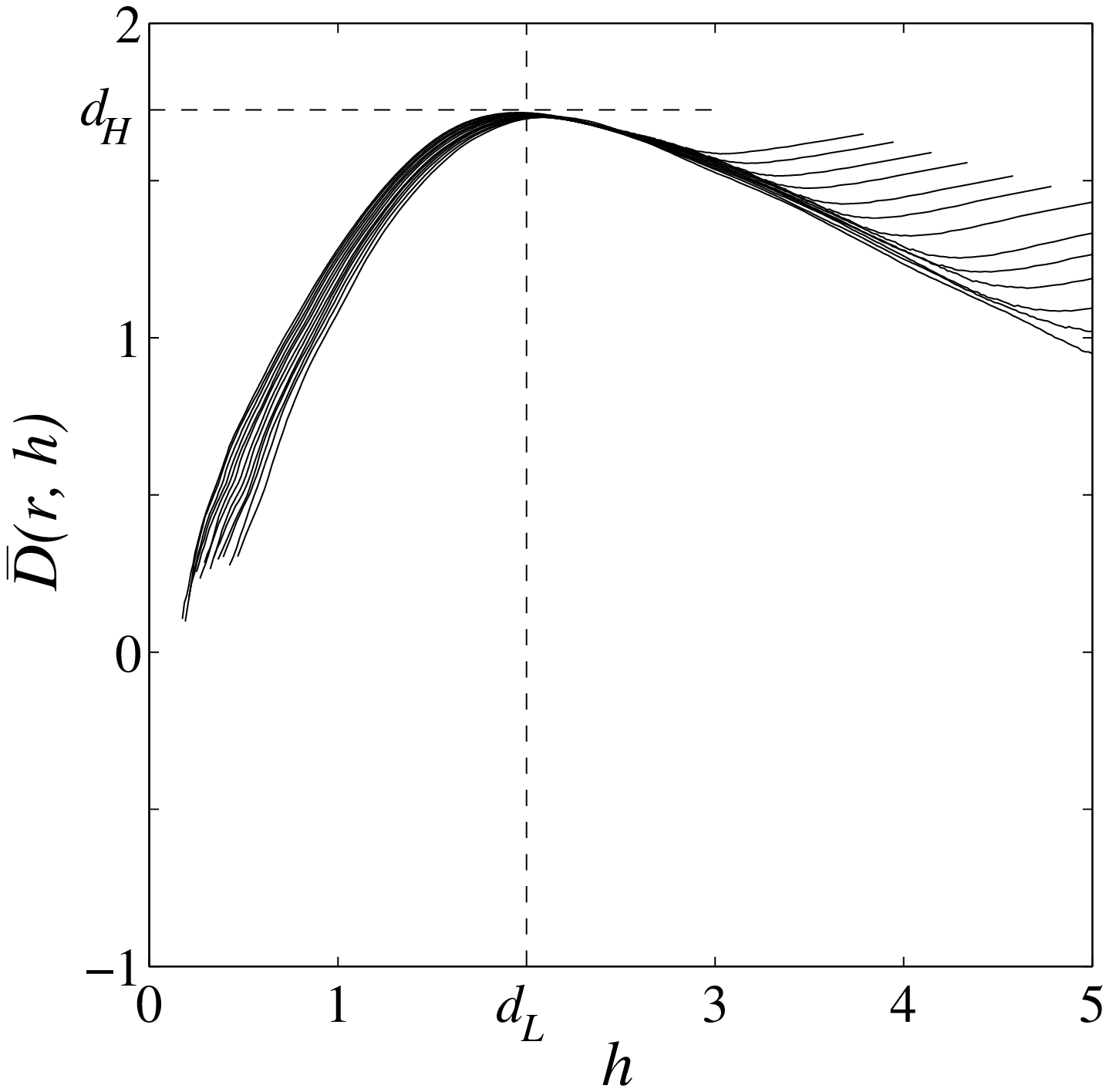}}}
  \caption{\label{fig:pdfm} Mass distribution for very heavy
    particles. The quantity $\overline{D}(r,h)$ defined by
    (\ref{eq:defDbar}) is plotted here for different non-dimensional
    radii $r/L$ ranging from $5\times 10^{-3}$ to $2\times 10^{-1}$
    and for two different values of the Stokes number as labelled. The
    density field is determined from the positions of $N=10^5$
    particles.  The curves approximately collapse on the function
    $\overline{D}_0(h)$ which measures mass fluctuation at those
    scales. It attains its maximum $d_H$ for $h=d_L$.}
\end{figure}
The discrepancy from a uniform distribution is even clearer when
measuring the fluctuations of the position-space mass distribution. 
In a way similar to what was done  for the multifractal analysis in 
phase space, we define
\begin{equation}
  \overline{D}(r,h) \equiv d - \frac{\ln p_r(h)}{\ln r},
  \label{eq:defDbar}
\end{equation}
where $p_r(h)$ is the PDF of $h=\ln \overline{m}_r/\ln r$.  Scaling of
the position-space mass $\overline{m}_r$ in the intermediate
asymptotics $\Stok \ll r/L \ll 1$ implies that in this range
$\overline{D}(r,h)$ depends only weakly on the ball radius $r$ and 
becomes thus essentially a function $\overline{D}_0(h)$.  
It is easily checked that $\zeta_n = \inf_{h}
(n\,h+d-\overline{D}_0(h))$. Together with (\ref{eq:legendre}) and
(\ref{eq:relexponents}) this implies that $\overline{D_0}(h) = d_H +
D_0(h)-2d$.  Thus, $\overline{D}_0(h)$ has  a maximum equal to $d_H$
for $h=d_L$.

In figure \ref{fig:pdfm}, the functions $\overline{D}(r,h)$
corresponding to two different Stokes number are superposed for
different values of the ball radius $r$ in the intermediate
asymptotics. These curves were obtained in the same setting as for
figure \ref{fig:exposants}, namely for infinitely heavy particles
($\beta=0$) and by time averaging over $10^4$ turnover times the
coarse-grained density obtained from the position of $10^5$ particles.
Pretty good scaling is observed since the curves almost collapse on a
same multifractal spectrum $\overline{D}_0(h)$. This implies in
particular that the multifractal formalism catches well the mass
distribution at scales within the intermediate range $\Stok \ll r/L
\ll 1$.

The Lyapunov dimensions associated to the two small values of $\Stok$
investigated here are almost equal, from which one might infer that
the mass distributions are the same.  In fact this is not the case:
the results of the numerical experiments presented here give two kinds
of evidence for a different behaviour.  First, for $\Stok \ll 1$, the
Hausdorff dimension of the attractor $d_H$ may not tend to $d$ as fast
as $d_L$ (i.e.\ it may approach $d$ slower than quadratically). This
implies that clustering may be stronger than that predicted from the
analysis of the Lyapunov exponents. Second, the large fluctuations of
the mass distribution (both at large and low values) which are
represented by the tails of the function $\overline{D}_0(h)$ differ
markedly for the two different low values of the Stokes numbers
considered.  At small Stokes numbers where the Stokes drag is very
strong, the inertial particles almost follow the motion of the
fluid. It is thus tempting to believe that the small effects due to
particle inertia can be easily understood and quantified.  As we have
seen, this is not really the case: the small Stokes number asymptotic
regime has a nontrivial clustering mechanism and there remains a
number of open questions.

\section{Concluding remarks}
\label{sec:conclusion}

We have shown that the clustering of inertial particles in spatially
smooth flow can be analysed using tools from dissipative dynamical
systems.  We thereby obtained evidence for the existence of a critical
Stokes number below which particles form dynamical fractal clusters in
position space, threshold whose existence for simple random flows was
confirmed in both two and three dimensions by numerical simulations.
We showed that the fluctuations in phase space of the singular mass
distribution are derivable from the dimension spectrum of a random
dynamical attractor.  It is however difficult to ``project this down''
to determining mass fluctuations merely in position space, because the
integration over velocities requires the knowledge of the fractal
properties of the attractor in the phase-space directions associated
to particle velocities.  This cannot be done without a fuller
understanding of the phase-space dynamics.  So far, we have obtained
only partial results in the asymptotics of low Stokes numbers, a
regime with many practical applications.

Let us mention some further questions arising in the low Stokes number
limit.  The evidence is that already at very low Stokes numbers, say
$10^{-4}$, particles form fractal clusters in position space.  More
precisely, our numerical results confirmed the prediction of
\cite{bff01} that the discrepancy from a uniform distribution measured
from the Lyapunov dimension goes quadratically to zero as $\Stok\to
0$.  A delicate issue is whether or not the fractal Hausdorff
dimension of the clusters tends to the dimension of the ambient space
that fast.  Obtaining the Hausdorff dimension numerically is difficult
and requires extensive computational resources.  Hence, a systematic
investigation of its dependence on the dynamical parameters is at the
moment out of reach. There is however an alternative approach to some
of the small Stokes number issues. As proposed by \cite{m87}, the
particle dynamics at low Stokes numbers in an incompressible flow can
be approximately described as an advection of simple passive tracers
by a synthetic flow comprising a small compressible component (see
also \S\ref{subsec:heuristics}).  In this approximation, we can then
work directly in position space without having first to understand the
full phase-space dynamics. In this simpler framework we can in
principle take advantage of a recently proposed relation between the
multifractal properties of the particle distribution (including the
Hausdorff dimension) and the fluctuations of the stretching rates
\cite[]{bgh03}.  For this we need to know the large-time statistical
properties associated to the local dynamics and, in particular, the
large deviations of the stretching rates from their limiting values,
the Lyapunov exponents.  For the moment such a determination has been
made analytically only for spatially smooth compressible variants of
the Kraichnan model, that is for velocity fields which are
delta-correlated in time \cite[see, e.g.,][]{fgv01}.  The problem is
that delta-correlated flow cannot be used for the low Stokes numbers
regime. Indeed, the expansion requires that the Stokes numbers be much
smaller than the (non-dimensional) correlation time of the carrier
velocity field.  A purely analytic investigation may thus be
difficult, but we can use a mixed theoretical and numerical strategy
for determining the multifractal properties in the low Stokes number
regime: use the aforementioned relation to express them in terms of
the large deviations of the stretching rates, the latter being
determined numerically. This should be much easier than direct
numerical determination of multifractal exponents by, for example,
box-counting methods.

Finally, an important and non-trivial extension concerns the
clustering properties of inertial particles at scales within the
inertial range of turbulence.  The dynamics of the flow at these
scales is close to Kolmogorov's 1941 theory; thus the velocity is not
smooth but, approximately, H\"older continuous of exponent $1/3$. As a
consequence, tracer particles separate in an explosive way, given by
Richardson's $t^{3/2}$ law. It is thus important to identify and to
understand the mechanisms leading to preferential concentrations of
particles at those scales where there is a strong competition between
clustering due to dissipative dynamics and the explosive Richardson
separation.  This is a challenging questions with serious
methodological difficulties.  On the one hand, numerical exploration
of the properties of mass distribution at those scales may require
very long time averages, as seen in the present study for smooth
flows.  Very long direct numerical simulations of three-dimensional
high Reynolds number flows are hardly feasible. It may hence be
necessary to focus on simpler flows, such as the synthetic turbulence
of the Kraichnan model, the two-dimensional inverse cascade or
large-eddy simulations.  On the other hand, an analytical approach to
this question can clearly not make use of standard dynamical systems
tools which assume smooth dynamics.  It thus seems of interest to
adapt to the inertial particle dynamics models and tools which have
been successfully applied in recent years to passive scalars advected
by non-smooth flow in Kolmogorov-like regimes.

\begin{acknowledgments}
I am deeply grateful to U.\ Frisch for many interesting discussions
and for his continuous support.  During this work, I have benefited
from stimulating and fruitful interactions with M.\ Cencini, K.\
Domelevo, G.\ Falkovich, P.\ Horvai, J.\ Mattingly and M.\ Stepanov.
This material is based upon work supported by the European Union under
contract HPRN-CT-2002-00300 and by the Indo-French Centre for the
Promotion of Advanced Research (IFCPAR 2404-2).  Numerical simulations
were performed in the framework of the SIVAM~II project at the
Observatoire de la C{\^{o}}te d'Azur.
\end{acknowledgments}

\bibliography{inertial}
\bibliographystyle{jfm}

\end{document}